# FinQAPT: Empowering Financial Decisions with End-to-End LLM-driven Question Answering Pipeline

Kuldeep Singh
singhku2@msu.edu
Michigan State University
East Lansing, Michigan, USA

Simerjot Kaur
simerjot.kaur@jpmchase.com
JPMorgan Chase & Co.
New York City, New York, USA

Charese Similey
charese.h.smiley@jpmchase.com
JPMorgan Chase & Co.
New York City, New York, USA

## Abstract

Financial decision-making hinges on the analysis of relevant information embedded in the enormous volume of documents in the financial domain. To address this challenge, we developed FinQAPT, an end-to-end pipeline that streamlines the identification of relevant financial reports based on a query, extracts pertinent context, and leverages Large Language Models (LLMs) to perform downstream tasks. To evaluate the pipeline, we experimented with various techniques to optimize the performance of each module using the FinQA dataset. We introduced a novel clustering-based negative sampling technique to enhance context extraction and a novel prompting method called Dynamic N-shot Prompting to boost the numerical question-answering capabilities of LLMs. At the module level, we achieved state-of-the-art accuracy on FinQA, attaining an accuracy of 80.6%. However, at the pipeline level, we observed decreased performance due to challenges in extracting relevant context from financial reports. We conducted a detailed error analysis of each module and the end-to-end pipeline, pinpointing specific challenges that must be addressed to develop a robust solution for handling complex financial tasks.

**ACM Reference Format:**
Kuldeep Singh, Simerjot Kaur, and Charese Similey. 2024. FinQAPT: Empowering Financial Decisions with End-to-End LLM-driven Question Answering Pipeline. In *5th ACM International Conference on AI in Finance (ICAIF '24), November 14–17, 2024, Brooklyn, NY, USA.* ACM, New York, NY, USA, 8 pages. https://doi.org/10.1145/3677052.3698682

## 1 Introduction

The financial domain is characterized by an enormous volume of documents, including various kinds of reports, news articles, regulatory filings, and research publications, containing a wealth of information critical for decision-making processes. For example, financial analysts frequently require rapid access to specific numerical data, such as the percentage change in a company's stock price following a major announcement. The ability to effectively extract context from these documents and manage downstream tasks, such as question answering (Q&A), reasoning, and search, is critically important for finance professionals. However, the sheer volume of data, coupled with specialized terminology, presents numerous challenges. Consequently, there is a pressing need for an automated end-to-end pipeline, that adeptly extracts and processes information from financial documents.

Recently, the advent of Large Language Models (LLMs), such as GPT-3.5 [1] and GPT-4 [17], has revolutionized the natural language processing (NLP) landscape, demonstrating exceptional performance in various applications, including reasoning, question answering, summarization, etc., with minimal supervision ([4], [6]). Their effectiveness is further enhanced by innovative prompting techniques like Chain of Thought [30], Self-Ask [20], React [31], and Self-Consistency [29]. However, these models often generate inaccurate or hallucinated information. We try to address these challenges in the financial domain by proposing an end-to-end pipeline that effectively extracts relevant context and integrates it with advanced prompting techniques to address downstream tasks in finance domain.

In particular, our proposed pipeline, FinQAPT, consists of three distinct modules: (a) FinPrimary, a primary retrieval module that interprets queries and identifies relevant financial reports as coarse-grained contexts; (b) FinContext, a retrieval module that extracts fine-grained contexts from relevant reports; and finally, (c) FinReader, a reader module that executes downstream tasks, specifically numerical reasoning task, utilizing LLMs. We addressed each module individually and developed innovative techniques to improve their individual performance on the FinQA dataset [7]. Our results indicate that although we achieved robust results at module level, the performance of the end-to-end pipeline was less optimal. The primary reason for this performance drop was the challenge of integrating relevant coarse-grained and fine-grained contexts from reports. This difficulty arises because financial reports have contexts dispersed across different pages and formats, leading to ambiguous financial contexts and hence inaccurate results. Our comprehensive error analysis of each module and the end-to-end pipeline highlights various challenges. Additionally, we achieved state-of-the-art results for FinReader module, highlighting the potential for improved financial analysis through innovative context integration techniques. Our main contributions can be summarized as follows:

---
[1]https://platform.openai.com/docs/models/gpt-3-5

- Developing FinQAPT, an end-to-end pipeline tailored for query analysis and answering in financial domain.
- Developing a novel clustering-based negative sampling technique to enhance context retrieval performance.
- Proposing Dynamic N-shot Prompting, a novel prompting technique that amplifies the numerical reasoning capabilities of LLMs. Our results achieve state-of-the-art (SOTA) accuracy of 80.6% on the FinReader module on FinQA dataset.
- Identifying and highlighting specific challenges at each level of the pipeline, hence providing insights to guide future research directions.

## 2 Related Work

Open-domain question answering has seen significant progress with methods like DrQA [5], R3 [28], and OpenBookQA [10, 11, 14, 16]. However, these methods focus on general knowledge sources and don't cater to the financial domain's unique challenges. In the financial domain, studies like DyRRen [13] and APOLLO [25] have made strides but don't offer a complete end-to-end solution for complex financial inquiries. They often rely on pretrained encoder models, which require extensive training data and fine-tuning.

The advent of Large Language Models (LLMs), such as GPT-4 [17] and PaLM2 [1], has significantly advanced NLP. However, they often produce hallucinated or contextually irrelevant information, limiting their effectiveness in domains like finance where accuracy is crucial. To address these limitations, significant work has been done on retrieval-augmented generation (RAG) ([2], [26], [21]) to provide relevant in-context to LLMs, reducing hallucinations. Additionally, several prompting techniques (Chain of Thought [30], Self-Ask [20], React [31], and Self-Consistency [29]) have been proposed. Despite these advancements, there remains a need for fine-tuned retriever models for domain-specific and task-specific tasks as shown by [8] and [9], as generalized retrievers struggle with specialized language, complex terms, and structure, such as in financial documents. Incorporating these techniques and models into a comprehensive financial domain pipeline remains a challenge.

Our FinQAPT pipeline builds upon the existing literature by developing an end-to-end solution specifically tailored to the financial domain. By integrating context extraction mechanisms and advanced prompting techniques, FinQAPT aims to harness the power of LLMs while minimizing the generation of hallucinated or irrelevant information. This approach sets FinQAPT apart from existing work and highlights its potential to address the unique challenges of financial Q&A and numerical reasoning tasks.

## 3 Methodology

Fig 1 illustrates our pipeline, which consists of three key modules: FinPrimary, FinContext, and FinReader, working sequentially in a retrieval-augmented generation fashion. The FinPrimary module decomposes the input query into simpler queries and identifies relevant pages from the S&P 500 Earnings Reports [32] to provide context for following stages. The FinContext module refines the context by identifying pertinent text segments and tables from the pages suggested by FinPrimary. Finally, the top five contexts retrieved, along with the original query, are passed to the FinReader module, which processes and generates the final answer. The subsequent subsections elaborate on the functionality of each module:

### 3.1 FinPrimary Module

The FinPrimary module selects coarse-grained information required to answer a given query, where coarse-grained refers to retrieving the relevant page from a document. Since queries can encompass multiple aspects, this module incorporates a query decomposition component. Consequently, the FinPrimary module operates in two distinct stages: query decomposition and retrieval of relevant pages or sections within the document.

*3.1.1 Query Decomposition.* We leveraged the advancements in LLMs, which have shown remarkable performance in various NLP tasks, including query decomposition ([19], [18], [3]). Specifically, we employed the 5-shot GPT-3.5 for decomposing our queries. An example is provided below,

```
Original Query : What is the percentage change in the
    fair value of the options for Apple from 2009 to
    2010 ?
Decomposed Query:
    1. What is the fair value of options for Apple in
        2009 ?
    2. What is the fair value of options for Apple in
        2010 ?
```

*3.1.2 Page Retrieval.* We extract the company name(s) and year(s) referenced in each decomposed query to select the relevant reports from the S&P dataset. Given that these reports can extend over hundreds of pages, it is important to identify the relevant pages that provide context for the specific query. Hence, for each individual query $Q$ and its associated earnings report comprising $N$ pages, we employ widely-adopted sparse-retrievers like TF-IDF [24], and BM25 [23], to retrieve the most relevant pages within the corresponding report.

### 3.2 FinContext Module

The goal of the FinContext module is to extract fine-grained information, such as multiple sentences or table rows, from the pages identified by FinPrimary for each query. Given a query ($Q$) and coarse-grained context ($C$), which includes both tables and text, our retriever extracts fine-grained context from $C$. To facilitate this process, we adopt a template akin to FinQA and DyRRen to transform tabular data into sentences. The template is similar to, *the **column name** of **row name** is **cell value**;*. This produces multiple sentences for a row of the table which are then concatenated together.

We treat this task as a binary classification problem. For a given query $Q$ and a context $C_i$ from $C$, we concatenate them and pass it through a pre-trained encoder model. The output embedding from the $[CLS]$ token is passed through a linear layer to compute a score $s_i$, indicating the relevance of $C_i$ to $Q$. Our retriever is trained on both positive and negative samples to ensure accurate $s_i$ computation.

$$s_i = Linear(Encoder_{output}(Q, C_i)) \tag{1}$$

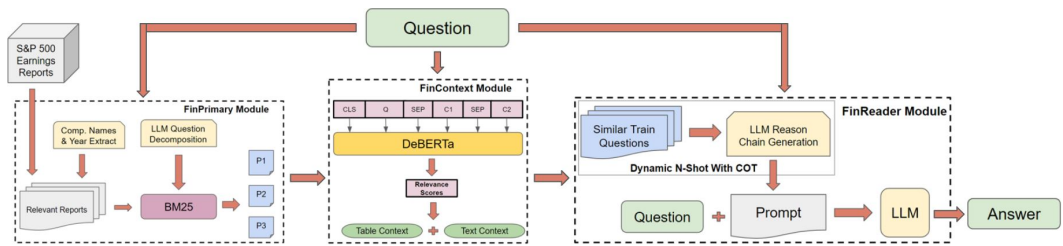

Figure 1: Architecture of FinQAPT pipeline consists of three modules: (a) FinPrimary for query interpretation and relevant report identification; (b) FinContext, for extracting fine-grained context from pages of reports; (c) FinReader, for downstream task with LLMs.

The performance of a binary classifier is significantly influenced by the quality of negative samples used during training. Recent studies like APOLLO have explored various negative sampling techniques, including random, self-mining, and Number Aware Negative Sampling. However, these techniques primarily focus on capturing numeric information in the context. In scenarios where the context is poorly constructed or where understanding the context requires textual information and complex inter-entity relationships, relying solely on number-centric negative sampling could lead to subpar performance.

Hence, a more comprehensive negative sampling technique that effectively captures both numeric and textual information is needed. In this study, we utilize **Clustering-based Negative Sampling** technique, inspired by [27], which leverages both sparse and dense representation methods for effective negative sample selection. We compute similarity scores between $Q$ and $C_i$ using TF-IDF, BM25, OpenAI [15], and ST (Sentence Transformer) [22] embeddings. These similarity scores are concatenated into a vector, representing each $C_i$ context in $C$.

$$v_i = [s^{TFIDF}(Q,C_i), s^{BM25}(Q,C_i), s^{OpenAI}(Q,C_i), s^{ST}(Q,C_i)] \quad (2)$$

Finally, we utilize K-nearest neighbors (KNN) on the constructed $v_i$ to identify the closest negative context to each positive once for every $Q$. These *hard* negative samples have similarity scores with the query that are comparable to those of the gold context. This approach of using both dense and sparse representation to select negative samples enhances the quality of training data, hence empowering the classifier to capture nuanced contextual details beyond mere numeric information.

### 3.3 FinReader Module

The FinReader module generates answers using the context extracted by the FinContext module, leveraging the impressive reasoning capabilities of LLMs. Prompting, a key element in LLM tasks, significantly influences the model's text generation, specifically the quality, relevance, and specificity. In this study, we introduce a novel prompting technique, **Dynamic N-Shot Prompting**, designed to enhance LLMs' numerical reasoning capabilities. It merges Chain of Thought (CoT) ([30]) prompting with our proposed technique of using training data.

*3.3.1 Dynamic Few-shot Prompt Generation.* Our method departs from traditional approaches where $N$-shot prompts are manually defined and remain static throughout the experiments. We use a dynamic approach that generates new $N$-shot prompts for each question in the test and dev datasets, with $N$ being the number of examples. This dynamic strategy outperforms static prompts by using semantically similar questions from the training data to construct prompts, ensuring more relevant and context-specific COT examples. It helps LLMs understand calculations of financial term using provided context and facilitates accurate computations. We leveraged the OpenAI-Ada-002 [2] embeddings, to identify semantically similar questions from training set using cosine similarity.

*3.3.2 Reasoning Chain Generation.* CoT prompting along with SC has proven effective in enhancing LLM performance, particularly in tasks that involve numerical reasoning or open-domain question answering. To seamlessly integrate CoT and SC with Dynamic N-Shot Prompt Generation, we generated reasoning chains for every question in the FinQA training set. This process was carried out using a 5-shot reasoning chain generation approach with GPT 3.5. The generation of reasoning chains was guided by *program steps*, already available for each question in FinQA. For example :

```
Question: What is the total operating expenses in
    2018 in millions?
Program steps: divide(9896, 23.6%) (provided in
    dataset)
The 5-shot reasoning chain generated using GPT-3.5 is:
```

1. The aircraft fuel expense in 2018 was $9896 million.
2. The percentage of total operating expenses attributed to aircraft fuel expense in 2018 was 23.6%.
3. To find the total operating expenses in 2018, we divide the aircraft fuel expense by the percentage of total operating expenses: $9896 / 23.6% = $41932.20339 million.
4. So the answer is 41932.20339.

This method allowed us to dynamically generate N-Shot Chain of Thought prompts for each question in the test and dev sets. Throughout the paper, any mention of "Dynamic N-Shot" refers to the combination of Dynamic Few-Shot with the Chain of Thought prompting method.

---
[2]https://openai.com/blog/new-and-improved-embedding-model

## 4 Experiments

### 4.1 Dataset

To test the individual modules and end-to-end pipeline, we conducted experiments using two datasets: the S&P 500 companies' earnings reports [32] and the FinQA dataset [7]. The FinQA dataset provides a sizable collection of carefully curated financial numerical reasoning Q&A and includes context information, indicating the specific report, i.e. the company ticker and year, and the page from which a context was taken in the S&P 500 earnings reports. It maybe noted that we leverage this metadata along with a dataset of S&P 500 reports to assess the performance of primary retrieval component of our pipeline.

#### 4.1.1 S&P 500 Earnings Reports. :

We downloaded the earnings reports of S&P 500 companies [3] spanning the years 1999 to 2019 used in the FinQA dataset. On an average an earnings report in the dataset consists of 300-500 pages, some reports even extend to even 700 pages. We extracted the text content from each page using pdfminer [4] and the table extraction was performed using Camelot [5].

#### 4.1.2 FinQA. :

FinQA dataset consists of a collection of 8,281 meticulously curated financial questions, annotated by financial experts. The dataset provided is split into training (70%), dev (15%) and test (15%) set. As mentioned above, FinQA offers metadata including the company ticker and year of S&P 500 report from which the context was extracted. However, some queries do not explicitly mention the company name and report year but implicitly use this information. To address this issue, we use the metadata to append the company name and year to the query wherever this information is missing.

### 4.2 Evaluation

#### 4.2.1 Evaluation Metrics.
To evaluate our retrieval modules, FinPrimary and FinContext, we employ Recall Top-N (Recall@N), which measures the percentage of correct positive predictions. In scenarios with multiple positive predictions per sample, we consider the first N predictions as positive. For FinReader module, we adopt a methodology akin to *PoT (Program of Thoughts Prompting)*. However, instead of solely employing *math.isclose* with a relative tolerance of *0.001* for answer comparison, we implement the following techniques to enhance the extraction and processing of generated outputs:

(1) Despite prompting the LLMs to produce JSON schema, the generated answers often necessitated additional text processing. Hence, we use a regex pattern to isolate the answer string from the LLM's generated output.
(2) If the extracted answer string contains *insufficient_context*, no further processing is performed.
(3) Ratios and percentages are converted to a standardized format. For instance, *90%* in the answer string or gold answers is converted to *0.90*. Ratio formats, such as *3/4*, are transformed to *0.75*.
(4) After normalizing the extracted answer string, it is directly compared with both the *answer* and *exe_answer* provided in the dataset. Correct matches are assigned a value of 1, while incorrect matches undergo an additional step.
(5) The normalized extracted answer string is compared with the *exe_answer* in the dataset using *math.isclose* with a tolerance of *0.01*. This approach is beneficial as LLMs may struggle with accurately computing high precision floats and large numbers.

#### 4.2.2 Evaluation Approaches.
We evaluated the performance of our system using two distinct approaches:

**Module-wise Evaluation:** This approach involved testing each module individually to assess its effectiveness in retrieving relevant information and answering queries. We first evaluated the pages extracted by FinPrimary module with the FinQA's metadata, i.e., the report and page from the S&P 500 Earnings reports used to create each query. We then tested the FinContext and FinReader modules together using the same metadata, allowing us to assess the retrieval of relevant contexts from the provided page. We then used these contexts, specifically Top-3, to evaluate the correct answer to the query.

**End-to-end Evaluation:** In this approach, we evaluated the entire pipeline, encompassing the FinPrimary, FinContext, and FinReader modules. We fed the relevant pages, specifically Top-8 from FinPrimary, into FinContext to extract refined contexts. These contexts, specifically Top-5, were then input into FinReader to generate the query's answer.

### 4.3 Training Hyperparameters

**Dense Retriever Models Hyperparameters:** The FinContext module's models were trained on *AWS g4dn.2xlarge* instances, employing a batch size of *4* and a learning rate of *2e-5* for a maximum of *10 epochs*. We utilized the PyTorch Lightning library for multi-GPU training, incorporating a *16-mixed precision* setting alongside distributed data parallel strategy.

**LLM Prompt Settings:** For the FinReader Module, utilizing OpenAI's API (for GPT-3.5-turbo and GPT-4 models), Google's VertexAI API (for PaLM2 model), and LLaMa2 model (using HuggingFace library), we maintained a fixed temperature of 0.2 during generation. All other parameters were kept as the provided default parameters in the API functions.

### 4.4 Results

#### 4.4.1 Module-wise Results. :

**FinPrimary** The evaluation results in Table 1 demonstrate that TF-IDF outperforms BM25. This is anticipated as TF-IDF excels in the financial domain by prioritizing unique financial terms and exact phrasing, which are critical for precise retrieval. Conversely, BM25 struggles to adapt to specialized language and structure of financial documents, making TF-IDF's term matching and importance weighting a better choice for capturing document relevance.

**FinContext** Table 2, present a comprehensive comparison of methods used in the FinContext module. The evaluation highlights that while pre-trained encoder models like OpenAI and Sentence

---
[3] https://www.annualreports.com/
[4] https://pypi.org/project/pdfminer/
[5] https://camelot-py.readthedocs.io/en/master/

Table 1: Evaluation results of FinPrimary @ Recall Top-3 and Top-5

| Experiment | Dev | | Test | |
|---|---|---|---|---|
| | R@3 | R@5 | R@3 | R@5 |
| **TF-IDF** | 81.3 | **89.4** | 83.1 | **89.2** |
| **BM25** | 78.8 | 85.7 | 79.9 | 85.5 |

Transformers perform comparably to sparse representation methods like BM25 and TF-IDF, they significantly lag behind fine-tuned methods. Our results demonstrate that fine-tuned models can boost R@3 and R@5 scores by over 20% compared to baseline methods. This underscores the limitations of generalized retrieval techniques in domain-specific Q&A and highlights the need for model fine-tuning to improve recall rates.

Table 2: Evaluation results of FinContext using various negative sampling strategies.

| Type | Model | Dev | | Test | |
|---|---|---|---|---|---|
| | | R@3 | R@5 | R@3 | R@5 |
| Baseline | TF-IDF | 72.1 | 72.7 | 81.9 | 82.6 |
| | BM25 | 73.1 | 74.0 | 82.9 | 83.0 |
| | OpenAI | 72.1 | 74.2 | 81.3 | 83.4 |
| | SentTrans | 71.9 | 73.7 | 80.7 | 83.0 |
| Random | BERT | 89.1 | 88.7 | 91.7 | 92.6 |
| | RoBERTa | 88.3 | 89.4 | 92.4 | 92.7 |
| | DeBERTa-v3 | 89.7 | 90.3 | 92.3 | 93.6 |
| Self-Mining | BERT | 86.1 | 87.0 | 89.2 | 89.4 |
| | RoBERTa | 86.8 | 89.1 | 89.7 | 91.7 |
| | DeBERTa-v3 | 87.9 | 88.2 | 90.0 | 91.1 |
| Cluster-based | BERT | 88.5 | 89.3 | 92.3 | 93.4 |
| | RoBERTa | 89.8 | 91.2 | 92.9 | 94.0 |
| | DeBERTa-v3 | 91.1 | **92.4** | 94.0 | **95.0** |

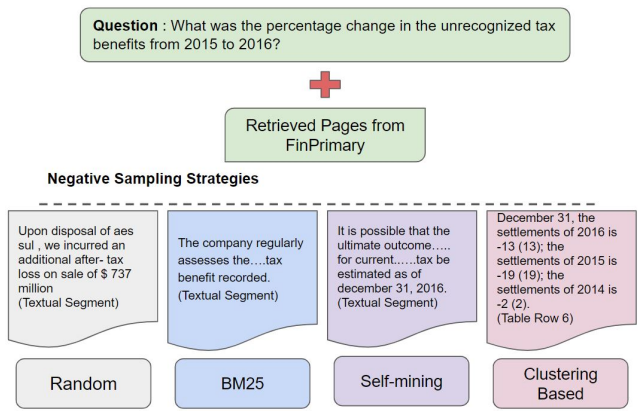

Figure 2: Comparison of different Negative Sampling strategies.

**FinReader** Table 3, present a comprehensive comparison of various prompting techniques on different LLMs, where the Recall @Top-3 contexts from FinContext are fed as relevant context along with the original query to LLMs. Our experiments on FinQA demonstrate the significant impact of our Dynamic $N$-shot prompting method in enhancing numerical reasoning capabilities across various LLMs, consistently achieving a 10%-15% improvement in accuracy. In our analysis, GPT-3.5 performs similar to PaLM2, while LLama2-7B lags behind, potentially due to its smaller parameter size. However, our approach significantly boosts the LLama2-7B model's performance. We achieve state-of-the-art (SOTA) results on both the dev and test sets of FinQA using our Dynamic N-shot prompting approach with GPT-4. As shown in Fig 3, our approach achieves 80.6% accuracy, surpassing the current SOTA model, APOLLO, with 71% and GPT-4 CoT results [12] with 78% accuracy.

Table 4: Results of FinQAPT pipeline, where Top-8 results from FinPrimary fed into FinContext, and Top-5 outputs from FinContext fed into FinReader to obtain accuracy on FinQA.

| Modules | Models | Dev | Test |
|---|---|---|---|
| **FinPrimary** | TF-IDF | 93.3 | 92.0 |
| **FinContext** | DeBERTa w/ C-Neg Sampling | 73.5 | 72.3 |
| **FinReader** | Dynamic 3-Shot GPT-4 | 63.1 | 62.0 |

*4.4.2 End-to-End Pipeline.* Table 4 displays the results of our entire pipeline using the best-performing method from the modular evaluation. The results show a performance drop in the FinContext module when used in a sequential pipeline, which subsequently impacts the FinReader module. This can be attributed to the structure of the FinQA dataset, which is based on a specific report page, with the query and context built solely around that page. Given the dataset's focus on numerical reasoning, the context for a numerical question can be dispersed across multiple report pages in various formats. Treating each page separately can lead to a loss of context, complicating the model's task of accurately answering complex queries.

This challenge is amplified when the query doesn't explicitly state the exact context. For instance, a query with phrases like *change in expenses* requires contextual information to identify the correct table. The expenses referenced could relate to total expenses, specific product expenses, etc. Consequently, when pages selected from FinPrimary are fed into FinContext, the performance of both FinContext and FinReader declines. We highlight specific challenges encountered during the development of an end-to-end pipeline for the financial domain in the error analysis section.

## 4.5 Error Analysis

*4.5.1 Module-wise Error Analysis.* We conducted thorough error analysis of each individual module in the pipeline:

**FinPrimary** :

REPETITIVE INFORMATION: Approximately 8% of questions seek information that is repeated in multiple sections of the financial report. For example, a question like *What were the operating expenses in 2006 in millions?* may require retrieving pages from the *Income*

Table 3: Evaluation results of FinReader on various prompting techniques across different LLMs. 3-Shots and 5-Shots here use static promtps whereas 3-Shots Dyn and 5-Shots Dyn use the dynamic n-shot prompt generation method from section 3.3.1

| Model | Zero Shot | | 3-Shots | | 5-Shots | | 3-Shots Dyn. | | 5-Shots Dyn. | |
|---|---|---|---|---|---|---|---|---|---|---|
| | Dev | Test | Dev | Test | Dev | Test | Dev | Test | Dev | Test |
| Llama2-7B | 10.0 | 8.2 | 13.3 | 14.8 | 12.8 | 14.8 | 39.6 | 35.5 | 40.2 | 37.4 |
| PaLM2 | 54.4 | 54.3 | 54.8 | 53.8 | 54.4 | 54.9 | 71.3 | 70.9 | 71.9 | 70.5 |
| GPT-3.5 | 56.4 | 52.7 | 55.0 | 53.1 | 56.1 | 53.7 | 68.9 | 67.5 | 72.3 | 68.8 |
| GPT-4 | 61.9 | 65.2 | 67.7 | 69.7 | 70.4 | 71.8 | **78.3** | **80.6** | 79.8 | 80.1 |

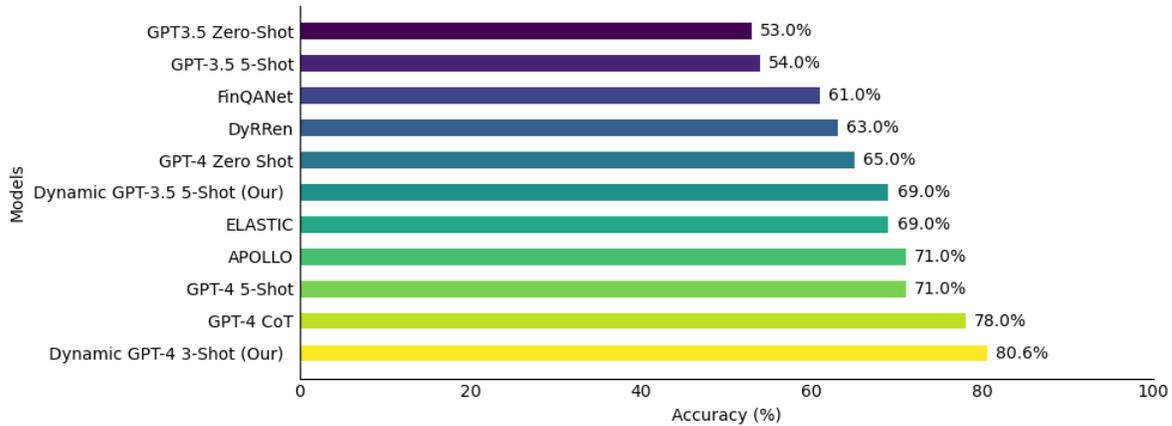

Figure 3: Comparison of SOTA models on FinQA test set, including LLMs. Our proposed Dynamic 3-shot prompting w/ GPT-4 approach achieves SOTA results with 80.6% accuracy.

*Statement, Financial Highlights, and Management's Discussion sections.* This repetition creates ambiguity for the model in identifying the most relevant context page.

QUESTION AMBIGUITY: Around 2% of questions are ambiguous and require contextual understanding to retrieve the correct information. For instance, a question like *What is the debt-to-asset ratio?* may require extracting pages with tables that include the ratio or pages with explanations of such financial terms. Therefore, providing explicit and clear questions can help reduce this ambiguity.

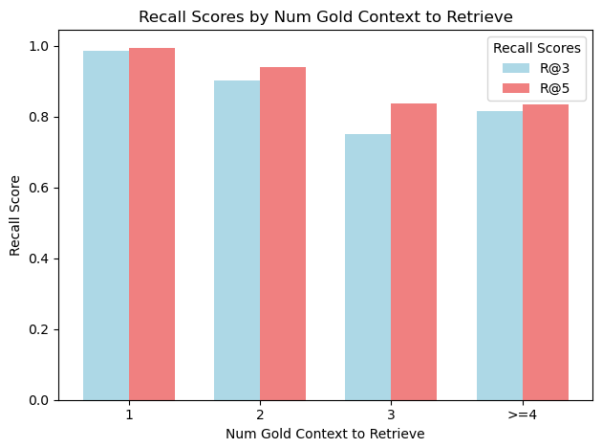

Figure 4: Increasing the number of Context to be retrieved impacts the FinContext Module.

**FinContext:**

AMOUNT OF CONTEXT RETRIEVAL: This category poses significant performance challenges in about 4% queries, Fig 4. Retrieval methods struggle when tasked with retrieving more than two sentences or table rows, with greater difficulties encountered when retrieving multiple sentences compared to multiple table rows.

COMPLEXITY OF CONTEXT RETRIEVAL: Accounting for about 7% of the queries, this aspect poses additional challenges for retrieval methods, especially when the context to be retrieved includes both text and table rows, Fig. 5. Specifically, issues arise when text and multiple table rows need to be retrieved, and the text, despite being irrelevant, is marked as gold context in the FinQA dataset.

**FinReader:** Our analysis reveals that while the PaLM model excels at identifying correct numbers for calculations, it often makes computational errors. Conversely, GPT-3.5 outperforms PaLM in calculation proficiency, but GPT-4 surpasses both in identifying correct numbers and precise computations, leading to significantly improved performance. Our investigation also found issues with question consistency, where the gold answer is expressed as a percentage for questions requiring a specific value. In other cases, disparities between the execution answer and the provided answer lead to misunderstandings of financial terms.

CONTEXTUAL COMPLEXITY: This category refers to ~5% of scenarios where LLMs struggle to identify the correct numbers for calculations or select values from a table for a different year than what is mentioned in the question. Inconsistencies in the context retrieved by the retriever modules also add to the complexity.

QUESTION COMPLEXITY: Encompassing ~6% of queries, this is where LLMs face challenges in understanding the question's intent, such as the presence of financial terms like *fair value* not found in the context or questions with specified formats. Discrepancies between the chosen method and the provided answer for projection-requiring questions can also result in incorrect responses.

Calculation Complexity: This aspect involves ~5% errors that arise when LLMs are tasked with calculating final answers that involve large digits or decimals, potentially leading to inaccuracies.

Insufficient Context: This category is particularly prominent in responses generated by GPT-3.5-turbo and GPT-4 compared to PaLM, accounting for ~5% of the queries. Our investigation revealed that GPT models tend to provide inadequate context when they cannot locate financial terms mentioned in the question within the provided context. Additionally, insufficiency arises when questions imply a projection into the future without explicit mention.

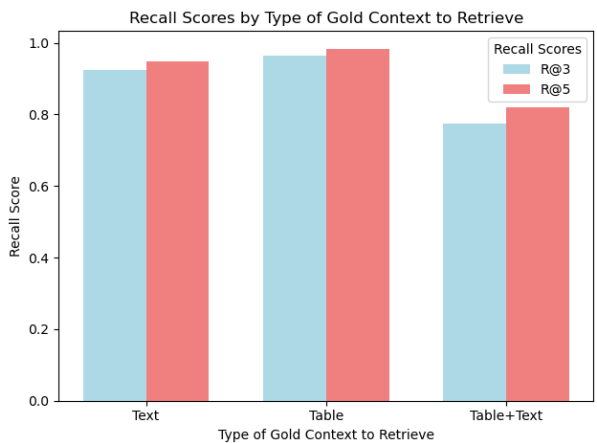

**Figure 5: Retrieving Table + Text context impacts the FinContext Module**

### 4.5.2 End-to-End Error Analysis.
We also conducted error Analysis on the End-to-End pipeline outputs and categorized the observed errors in below categories:

Contextual Reasoning: The FinContext module faces challenges in extracting relevant context from multiple report pages, impacting the FinReader module's performance. Additionally, some queries are inherently complex, requiring context from more than one report page, further complicating the model's ability to accurately answer such questions.

Noisy Information: Similar information appearing in multiple places within a financial report can mislead the model, resulting in inaccurate predictions. Enhancing the preprocessing of financial reports to eliminate such noise can improve the model's performance.

Implicit Context: The dataset's construction presents a challenge as some questions implicitly expect the model to understand the context, even when it's not explicitly stated in the question. This makes it difficult for the model to identify the relevant information required to accurately answer the question. Therefore, refining the dataset by providing explicit and clear questions can enhance the model's performance.

## 5 Conclusion

In this study, we underscore the challenges of constructing an end-to-end pipeline for handling complex financial tasks and integrating relevant context from financial reports. Despite achieving SOTA results for the FinReader module, the end-to-end pipeline's performance declined due to the loss of pertinent financial context. Our work offers valuable insights into enhancing the performance of each module and the end-to-end pipeline for financial analysis.

Our future work will focus on developing advanced techniques for integrating relevant context from multiple pages of financial reports to enhance the FinContext module's performance, and exploring alternative models to increase the pipeline's accuracy and robustness. Furthermore, our work emphasizes the need for more comprehensive datasets that encapsulate the complexity and nuances of financial analysis tasks.

## 6 Disclaimer